\documentclass[preprint,sort&compress,12pt]{elsarticle}

\usepackage{amssymb}
\usepackage{natbib}
\usepackage{amsmath,amsthm,bm,mathrsfs}
\usepackage{mathtools} 
\usepackage{epsfig}
\usepackage[colorlinks]{hyperref}
\usepackage{relsize}
\usepackage{xcolor}
\usepackage{url}
\usepackage{mathpazo}
\usepackage{pdflscape}
\usepackage{subcaption}
\usepackage{booktabs}
\usepackage{makecell}
\usepackage[normalem]{ulem}
\useunder{\uline}{\ul}{}

\journal{Physica A. \hspace{0.5cm} https://doi.org/10.1016/j.physa.2019.04.147}

\begin{document}

\begin{frontmatter}
	\title{Nonlinear dependencies on Brazilian equity network from mutual information minimum spanning trees}
	
	\author{A. Q. Barbi} 
	\ead{alex.barbi@usp.br}
	\author{G. A. Prataviera}
	\ead{prataviera@usp.br}
	\address{Departamento de Administra\c{c}\~ao, FEA-RP, Universidade de S\~{a}o Paulo, 14040-905, Ribeir\~{a}o Preto, SP, Brazil}

\begin{abstract} 
	\small
	Mutual information minimum spanning trees are used to explore nonlinear dependencies on Brazilian equity network by comparing the periods from June/01/2015 to January/26/2016, in which Brazil was under the government of President Dilma Rousseff, and from January/27/2016 to September/08/2016 which includes the government transition from President Dilma Rousseff to President Michel Temer. Minimum spanning trees from mutual information and linear correlation between stocks returns were obtained and compared. Mutual information minimum spanning trees present higher degree of robustness and evidence of power law tail in the weighted degree distribution, indicating more risk in terms of volatility transmission than it is expected by the analysis based on linear correlation. In particular, a remarkable increase of stock returns nonlinear dependencies indicates that the period including the government transition is more risky in terms of volatility transmission network structure. Also, we found evidence of network structure and stock performance relationship. Besides, those results emphasize the usefulness of mutual information network analysis for identification of Financial Markets features due to nonlinear dependencies.
\end{abstract}

\end{frontmatter}

\section{Introduction}
\label{intro}
Financial Markets are complex systems whose structure and behavior are strongly dependent on their components interrelations. In particular, network theory has contributed to characterize and understand the behavior of financial markets. Previous studies indicate that market network structure may contain useful information to characterize and even to predict the market behavior such as the occurrence of a financial crisis. An interesting method to analyse such structures started with the seminal work of ref. \cite{mantegna_1999-9u} using minimum spanning trees (MST). The MST is particularly suitable for extracting the most important information when a large number of linkages is under consideration. However, most of these studies are based on linear dependencies given by Pearson's linear correlation coefficient \cite{miccich_2003-4w, coelho_2007-Rg, tabak_2009-KZ, tabak_2010-Ab, gilmore_2010-8b, tab1, zhang_2011-iD, tab2, tab3, tab4, heiberger_2014-lp}. Thus, it is important to introduce measures of nonlinear dependence between components of a network. In particular, mutual information is a promising alternative to Pearson's coefficient as a measure of nonlinear dependence \cite{fraser_1986-gH, kraskov_2004--8, kinney_2014-rB}. Actually, the combination of information theory and network analysis has already proved useful for research on financial markets \cite {dionisio_2004-Fg, yang_2014-7H, fiedor_2014-m1}. Moreover, few studies \cite{kaya_2016} systematically compares the networks generated through Pearson's correlation with those generated through mutual information.

The Brazilian equity market is one of the most important in Latin America and of emerging markets. It has a significant number of worldwide big companies, like Ita\'u Bank (42nd), Bradesco Bank (67th), Vale do Rio Doce (2nd mining company) and Petrobr\'as (13rd oil \& gas company), according to the latest Forbes ranking. Additionally, Brazil is currently the 9th largest economy in the world, according to International Monetary Fund. However, the country has been the target of a series of corruption and money laundering investigations, notably the largest operation of this kind in the world, the so called "Lava Jato" (Car Wash) operation. This political turmoil has brought great volatility to financial market in recent years, and in particular during the President Dilma Rousseff's impeachment event in 2016. Actually, Dilma Rousseff was the president for the first time from 2011 to 2014, and was re-elected for a second time starting in 2015. Since then a political crisis has evolved towards her suspension in May-2016, and culminating in her definitive removal from office in August-2016. Michell Temer replaced Dilma Rousseff temporarily from May-2016 to August-2016 when he was definitely proclaimed president just after the official impeachment, and he remained president until December-2018. Thus, understanding the economic and political role in the financial network is important, and concerning the Brazilian market, the few network-based studies are restricted to MST based on linear dependencies using Pearson's correlation coefficient \cite{tabak_2009-KZ, tabak_2010-Ab, tab1}. So, we consider important to extend this kind of analysis taking into account asset's nonlinear interactions.

In this work we consider the role of nonlinear dependencies on the Brazilian equity network by comparing two consecutive periods selected from the in set of historical events above described. The first one from June/01/2015 to January/26/2016, in which Brazil was under the government of President Dilma Rousseff, and the second one from January/27/2016 to September/08/2016, corresponding to more critical period including the government transition from President Dilma Rousseff to President Michel Temer. We are interested in those periods because they are very dissimilar even during a short time window. While during the first period, stocks values went down more than 40\%, just six months after, stocks valued up around 50\%. Then is expected that in periods of high volatility nonlinear dependencies may show up predominant. We investigated such turbulence analysing networks structures obtained from Bovespa's high frequency stock returns. The main objective is to characterize both periods comparing the networks obtained from mutual information and Pearson linear correlation. Our results suggest that the networks obtained from mutual information present a more interconnected overall dependence, with stronger volatility transmission structure, notably during the second period. Finally, the analysis of the network in the financial market via mutual information brings benefits to the investment decision making, particularly with regard to the analysis of the network structure and stock's performance.

The article is organized as follows: in Section (\ref{sec2}) we discuss the methodology to obtain and analyse mutual information minimum spanning trees. In Section (\ref{sec3}) we present the data, and Section (\ref{sec4}) shows the results by contrasting the networks obtained from mutual information with those obtained with Pearson's linear correlation. Finally, in Section (\ref{sec5}) we present the conclusion.

\section{Methodology}
\label{sec2}
\subsection{Nonlinear dependence and Minimum Spanning Tree}
Here we are interested in analyse dependencies between stock price returns. The return of a given asset at time $t$ is defined by  
\begin{equation}
r_i (t) = ln \left(\frac{p_{i} (t)}{p_{i}(t-1)}\right),
\label{ret_log}
\end{equation}
where $p_i(t)$ is the price of stock $i$ at time $t$. To quantify the degree of dependence between assets we use a parameter based on mutual information. The mutual information between two stock returns $r_{i}$ and $r_{j}$ with joint probability density $f(r_{i},{r_j})$ is defined as \cite{shannon_1948-5A, cover_2005-yE}

\begin{equation}
I_{ij}= \int\int f(r_{i},r_{j})  \ln{\left(\frac{f(r_{i},r_{j}) }{f(r_{i})f(r_{j})}\right)} dr_{i} \, dr_{j},
\label{MI}
\end{equation}
where $f(r_{i})$ and $f(r_{j})$ are the marginal densities of $r_i$ and $r_j$, respectively. 
The MI is zero if the random variables $r_i$ and $r_j$ are statistically independent, i.e., if the joint density factorizes as $f(r_{i},r_{j})= f(r_{i})f(r_{j})$. 
Also, it is well known that statistical independence imply the vanishing of Pearson's correlation coefficient $\rho_{i,j}$ defined by
\begin{equation}
\rho_{ij}=\frac{\langle r_i r_j\rangle-\langle r_i\rangle \langle r_j\rangle}{\sqrt{(\langle r_{i}^{2}\rangle-\langle r_{i}\rangle^2) \, (\langle r_{j}^{2}\rangle-\langle r_{j}\rangle^2)}},
\end{equation}
which is easily estimated from empirical data, and where $\langle.\rangle$ stands for statistical mean. However, the vanishing of correlation does not in general imply that $r_{i}$ and $r_{j}$ are independent. Therefore, MI is a more general measure of dependence that goes beyond the usual linear relationship given by Pearson's correlation.  

On the other hand, Pearson's correlation and MI are not directly comparable since the first one varies in the interval $\left[-1,1\right]$ while the second one may assume values in the interval $\left[0, \infty \right)$. Then, they should be scaled over the same range. In addition, the degree of linear dependence given by Pearson's correlation is better associated with its absolute or its square value. In this paper we use a dependence coefficient whose definition is based on the fact that the MI for a bivariate normal density is given by \cite{cover_2005-yE}
\begin{equation}
I_{ij}=-\frac{1}{2}\ln{\left(1-\rho_{ij}^2\right)}.
\label{normal}
\end{equation}
Eq. (\ref{normal}) shows that correlation implies dependence for a bivariate normal distribution . Now, the inversion of Eq. (\ref{normal}) allows to define the following global coefficient of dependence \cite{joe_1989-GX,dionisio_2004-Fg}
\begin{equation}
\lambda_{ij}= \sqrt{1-e^{-2I_{ij}}},
\label{global}
\end{equation}
which is limited to the interval $[0,1]$, and can be compared with the linear correlation absolute value. In addition, a value of $\lambda_{ij}>\rho_{ij}\approx 0$ indicates nonlinear dependence.

With the global dependence coefficient we can define a distance function 
\begin{equation}
d_{ij} \, (global) = 1-\lambda_{ij},
\label{d_lamb}
\end{equation}
which can be directly compared with a distance based on Pearson's correlation coefficient 
\begin{equation}
d_{ij} \, (linear) = 1 -\mid  \rho_{ij} \mid.
\label{dist_pearson}
\end{equation}
The distances defined by Eqs. \eqref{d_lamb} and \eqref{dist_pearson} will be used to obtain the network minimum spanning tree (MST). The MST is a simply connected graph that connects all $N$ nodes with $N - 1$ edges such that the sum of all edge distances is a minimum, thus reducing the information contained in $N(N-1)/2$ dependence coefficients to $(N-1)$ edges. Kruskal's algorithm \cite{kruskal_1956-0c} is one of the main algorithms for searching a MST. It adds new weaker links until there are no new additions without making a cycle. This process finishes when the graph is fully connected.

Furthermore, to obtain the MI and the coefficient of dependence in Eq. \ref{global}, we need to estimate the empirical density functions $f(r_{i},r_{j})$, $f(r_{i})$, and $f(r_{j})$ from data. Here we use the non-parametric method of kernel density estimation \cite{silverman_1986-VY, scott_2015-Ra, moon_1995-4U}. For a $d$-dimensional set of variables $\textbf{x}=(r_{1},\cdots,r_{d})^{{T}}$ and the given data set $\{ \bf{r}_{1},\cdots,\bf{r}_{n}\}$ whose density is to be estimate, the multivariate kernel density estimator with kernel $K$ and window widths $h_{1},\cdots ,h_{d}$ is defined as 
\begin{equation}
f(\textbf{x}) = \frac{1}{nh_{1}\cdots h_{d}} \, \mathlarger \sum_{i=1}^{n} \, \left[\mathlarger \prod_{j=1}^{d}K\left( \frac{r_{j}-r_{ij}}{h_{j}}\right)\right],
\end{equation}
where the kernel function satisfies $\int_{d}K(\textbf{x}) d\textbf{x}=1$. In this work we use the Gaussian kernel, which for a radially symmetric standard multivariate normal distribution is written as
\begin{equation}
K(\textbf{x}) = \frac{e^{-\frac{1}{2}\textbf{x}^{{T}}\textbf{x}}}{(2\pi)^{d/2}}.
\end{equation}
Besides, for the window bandwidth we use the criteria that minimizes the density mean integrated square error and whose optimal value for a multivariate Gaussian kernel is given by \cite{scott_2015-Ra,moon_1995-4U} 
\begin{equation}
h_{i}=\left( \frac{4}{d+2}\right)^{\frac{1}{d+4}} \hat{\sigma}_{i} \, n^{-\frac{1}{(d+4)}},
\label{h_opt_m}
\end{equation}
where $\hat{\sigma}_{i}$ is the sample standard deviation of variable $i$, and $d=2$ for a bivariate kernel.

In addition, the statistical significance for the estimated correlation and mutual information was taken into account as follows: we set the significance level $\alpha=0.01$ and assign zero for dependencies (correlation or mutual information) with $p$-value $\geq \alpha $. Statistical significance for mutual information was assessed by a Chi-square test of independence, while for correlation a $t$-test was carried out \cite{cellucci,rice2006,raval2013introduction}.

\subsection{Network measures}
We also calculate a variety of network measures to characterize the MSTs. An important aspect of financial networks are their degree distribution. For example, a power law distribution is typical of scale free networks where hubs are much more connected then what is expected by Poisson distribution of random networks. This lack of scale can be analysed by the scaling parameter $\alpha$ of a power law distribution, whose calculation is given by the maximum likelihood estimator \cite{barabasi_2016-uP}
\begin{equation}
\alpha = 1 + n \, \left[\mathlarger \sum_{i=1}^{n} \, ln \left(\frac{x_i}{x_{min}}\right)\right]^{-1},
\label{power}
\end{equation}
where $x_i$ are the observed degree values, $x_i \geq x_{min}$, and $x_{min}$ is a cut-off parameter that can be estimated using Kolmogorov-Smirnov statistic \cite{clauset_2009-26}. For this purpose, the R poweRlaw package \cite{powerR} is employed. Since we are interested in the strength of the dependencies, we use the weighted degree distribution in our analysis. This means that a given degree will be weighted by the values of $\lambda_{ij}$ or $|\rho_{ij}|$.

To estimate network's financial systemic risk, we can determine its robustness coefficient, given by \cite{barabasi_2016-uP}
\begin{equation}
R = 1 - \frac{1} {\frac{\langle k^2 \rangle}{\langle k \rangle} - 1},
\label{mollow3}
\end{equation}
where $\langle k \rangle$ and $\langle k^2 \rangle$ are the degree distribution first and second moments, respectively. The robustness coefficient represents the fraction of nodes in a network that we need to remove to lose its component structure. Here, we use this metric to evaluate the network's concentration structure. The bigger the nodes that are very highly connected (a hub), the higher the network's robustness coefficient, until it approximates to 1. 

In order to measure hubs strength individually, we use the node's eigenvector centrality. This parameter indicates that a node is important if it is connected to other nodes that are also important. So, the centrality $x_{i}$ of vertex $i$ is defined by \cite{newman_2010-jW}
\begin{equation}
x_i = \kappa_1^{-1} \mathlarger \sum_j {A_{ij} {x_j}},
\label{centralidade2_1}
\end{equation}
where for us $A_{ij}=\lambda_{ij}$ (or $|\rho_{ij}|$) are the entries of the MST weighted adjacency matrix $A$, $\kappa_1$ is the largest eigenvalue of $A$, and $x_{j}$ is the centrality of neighbors of $i$. 

Those measures are useful to identify volatility transmission in financial networks. Volatility transmission means that once a hub gets affected by certain volatility, it spreads that instability through its connections, possibly causing a financial instability. This same logic can be used for a variety of systems, like the power grid network. For example, on August 10, 1996, Oregon, a line carrying 1,300 megawatts sagged close to a tree and snapped, while its current was automatically shifted to two lower-voltage lines. However, these also failed. This day, a full blackout happened in 11 US states. In this example, that line (a hub) had a significant energy that could not be tolerated by the other lines if it fails, possibly causing a cascade effect. In financial networks, an important asset (hub) can also be affected by a certain volatility, such as a news about an accounting fraud in a big bank. Then, quickly many others might feel the impact of the news, especially if they have with it a strong dependence. Thus, a high value of robustness coefficient means that the network has a strong interconnected structure, that is, a significant overall dependence among assets. More, that it has a very significant hub. In this way, there's more probability that a great turmoil affects more and more stocks, like in a domino effect \cite{dobson_2007-mp}. 

\section{Data}
\label{sec3}
The data set for this paper comes from the Bovespa Index (Ibovespa), which comprises a portfolio of the main Brazilian assets, and is the most important index of average performance for the Brazilian stock market prices. The share of each stock in the Ibovespa portfolio is directly related to its representativeness in terms of the number of trades and financial volume. This is given by the Stock Negotiability Index (SNI)

\begin{equation}
SNI = \sqrt{\frac{n_i}{N} \times \frac{v_i}{V}},
\label{ibovespa_formula}
\end{equation}
where $ n_i $ is the number of deals with the stock $ i $, $ N $ is the total number of trades, $ v_i $ is the financial volume generated by the trades with the $ i $ share, and $ V $ is the total volume traded considering the spot market for all variables. We calculated the SNI from a window of 5 years (2012-2016) and selected the first 20\% of the total shares with the highest negotiability. This sample gets almost 80\% of the SNI, whose values were obtained from Economatica \cite{economatica}. 

Our analysis is restricted to data from two periods that we consider relevant. The first period, from June/01/2015 to January/26/2016, was under the government of President Dilma Rousseff, with 159 trading sessions and index return of -42\%, where we collected 3888 stock returns, given every 15 minutes. The second one, a period from January/27/2016 to September/08/2016,  corresponding to the transition from Dilma Rousseff to Michel Temer, with 160 trading sessions and index return of 50\%, where we obtained 3969 returns, given every 15 minutes. Data were collected from Bovespa ftp site \cite{bovespa}. It contains high frequency trading data for various Brazilian financial instruments. For this purpose, we used the GetHFData package from R software \cite{HFT}.

All statistical and network analysis obtained in this paper were carried out in \textbf{R} software, while Gephi was useful for network illustration design.

\section{Results and discussion}\label{sec4}

\begin{figure*}[!htbp]
\centering
\begin{subfigure}[t]{.75\textwidth}
	\centering
	\includegraphics[width=\linewidth]{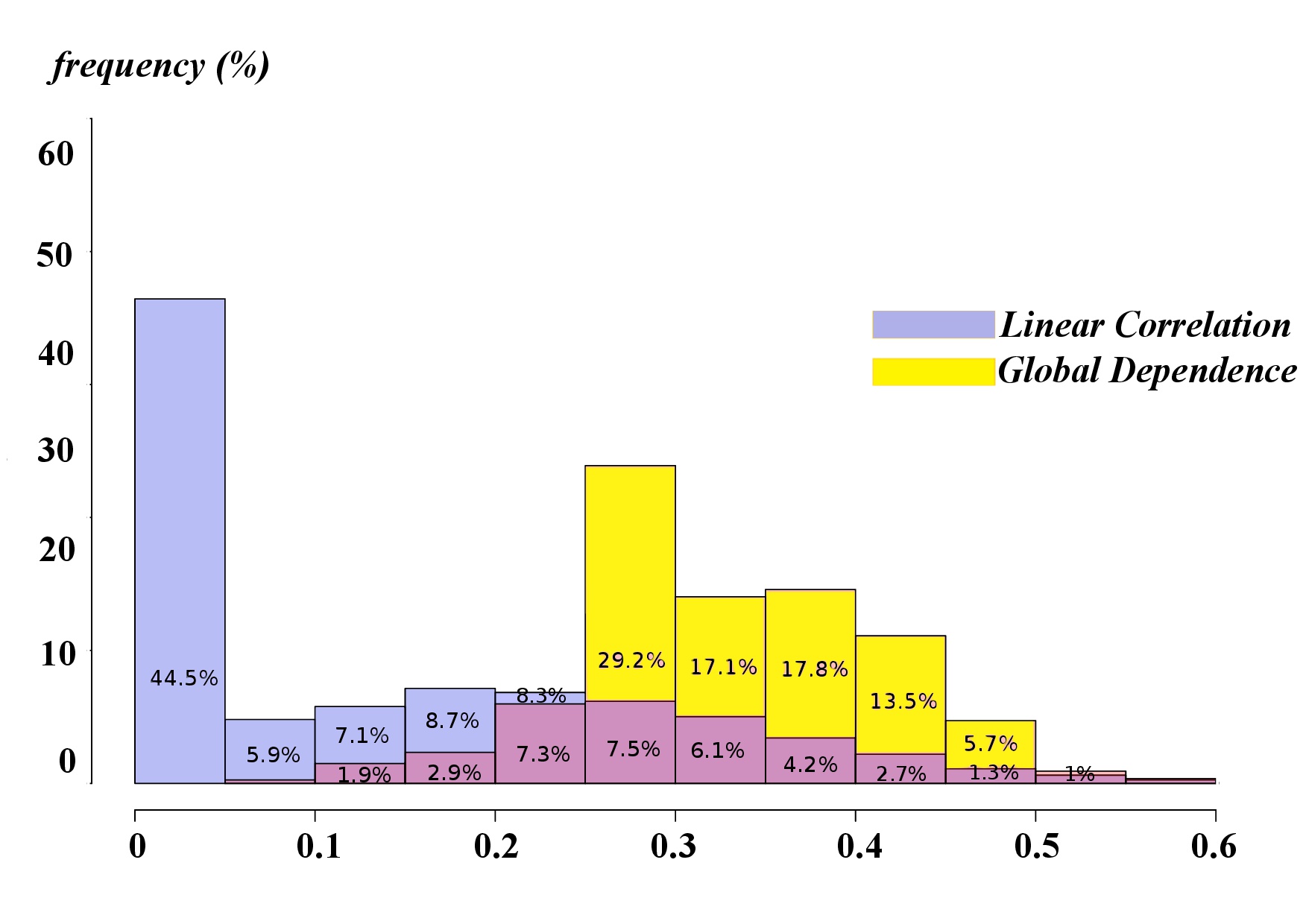}\vspace{-.5cm}
	\caption{DR period}
	\label{fig3a}
\end{subfigure}
\hfill
\begin{subfigure}[t]{.75\textwidth}
	\centering
	\includegraphics[width=\linewidth]{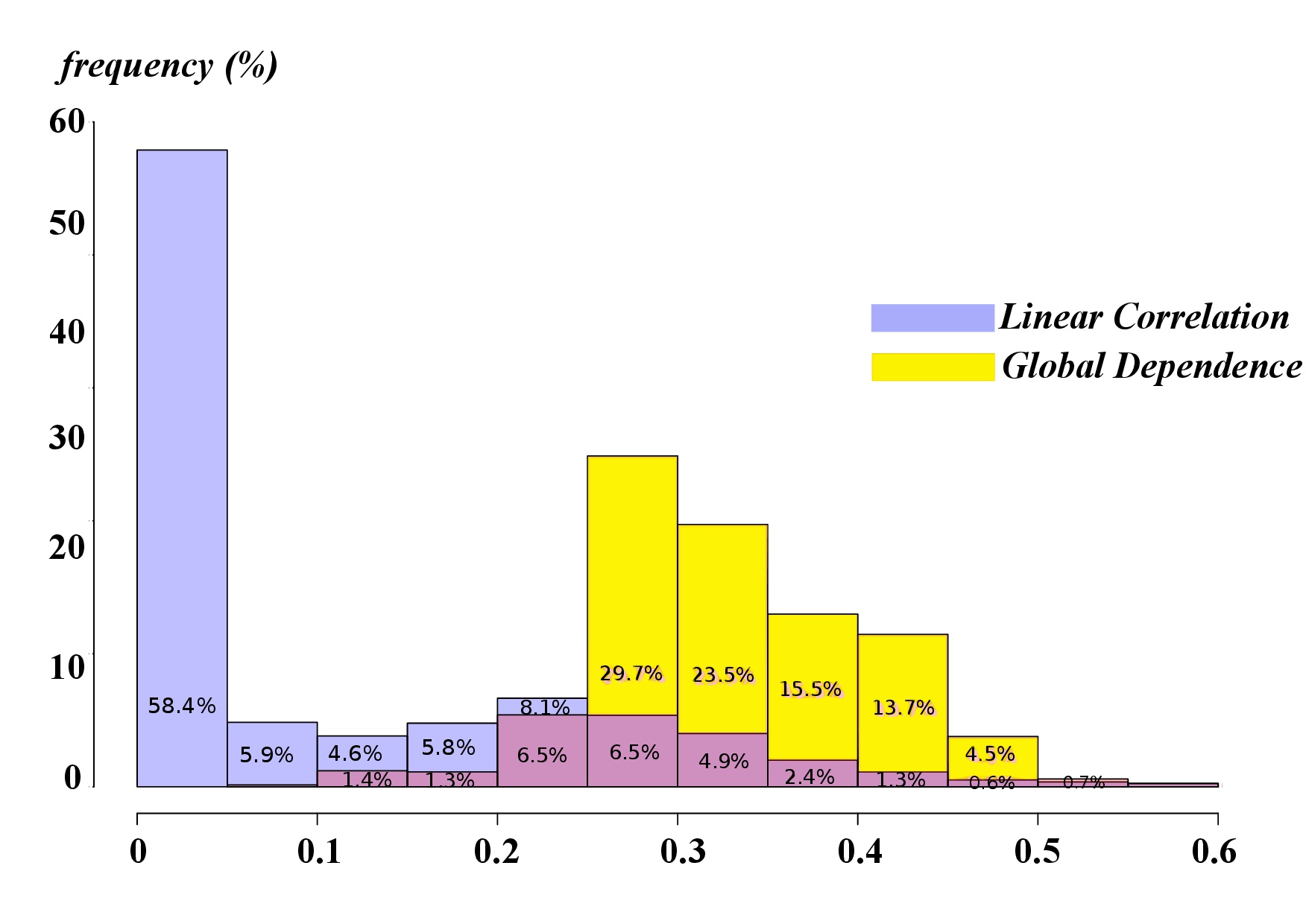}\vspace{-.5cm}
	\caption{MT period}
	\label{fig3b}
\end{subfigure}%
\caption{Distribution of linear (blue) and global dependence (yellow) in absolute values. Its intersection is in purple. (a) Dilma Rousseff's period; (b) Michel Temer's period. }
\label{fig3}
\end{figure*}

\begin{figure}[!htbp]\vspace{-3cm}
\begin{center}
	\includegraphics[scale=0.7]{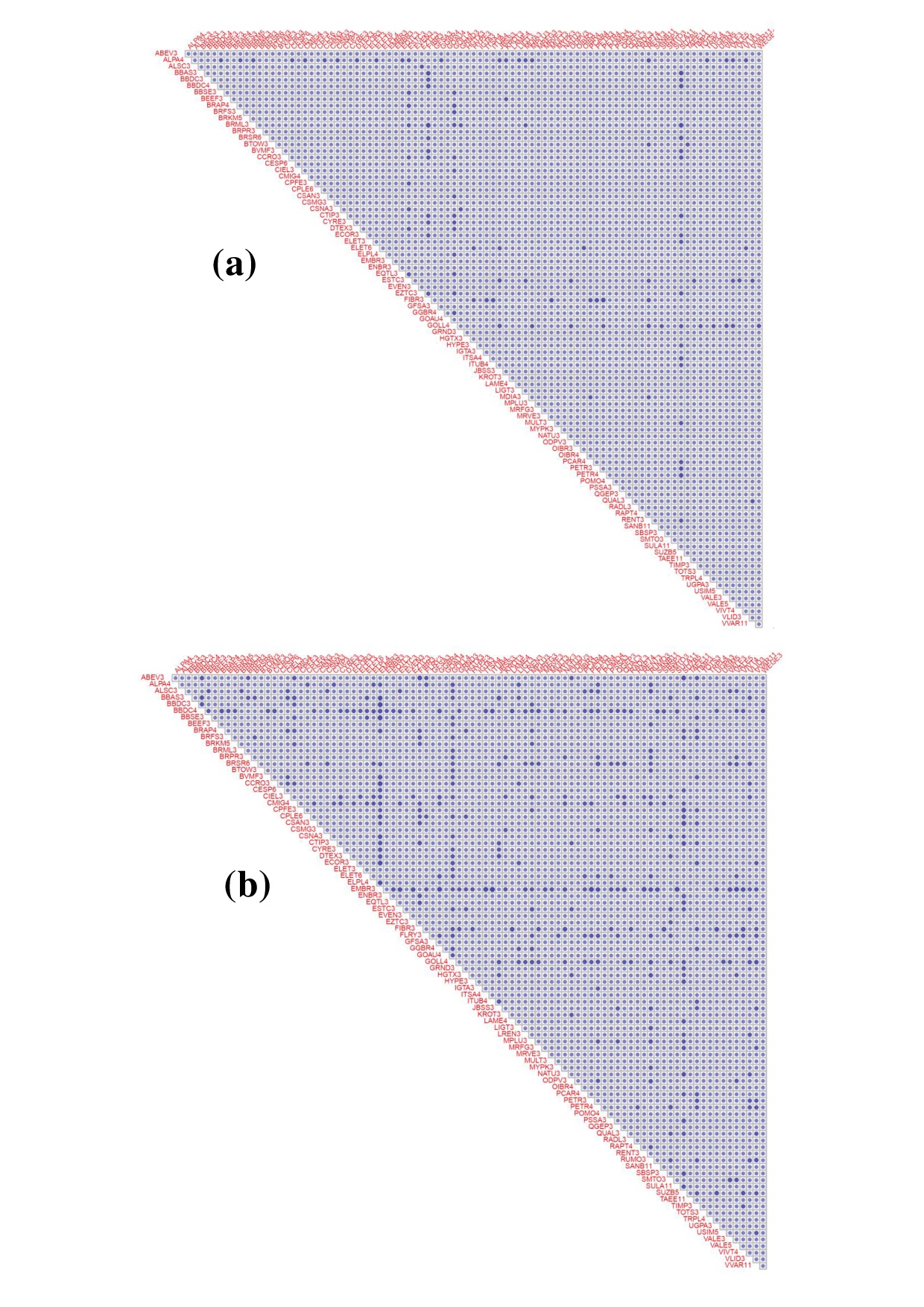}
\end{center}
\caption{Pictorial representation of the symmetric matrix whose entries are given by $( \lambda_{ij}-|\rho_{ij}|)$. (a) Dilma Rousseff's period; (b) Michel Temer's period. Dark dots show differences of at least 0.30.}
\label{fig4}
\end{figure}

To emphasize the role of nonlinear dependencies, we plot in Figs. \ref{fig3a} and \ref{fig3b} the frequency distribution of the absolute values of linear correlation and global dependence coefficient for the two periods. From now on we call the  first period as Dilma Rousseff (DR) period, while the second one including the transition from Dilma Roussef to Michel Temer will simply be called Michel Temer (MT) period. For MT period, about 82\% of the global dependence distribution is concentrated from 0.25 to 0.45, whereas for DR period this value is around 77\%. In addition, there is a higher proportion of linear correlation values between 0 and 0.1 in MT period than in DR period, which means that the overall difference of mutual information and linear correlation is bigger in MT than in DR period. This conclusion is reinforced by Figs. \ref{fig4}a-b, where we show a pictorial representation of the symmetric matrix whose entries are given by the difference between the global dependence coefficient and the absolute value of linear correlation for both periods, and whose entries are represented by color intensity, so that dark points corresponds to a difference of at least $0.3$. The higher concentration of dark points in Fig. \ref{fig4}b shows that this period is characterized by an increasing asset's nonlinear dependence. In fact, the linear correlation structure in MT period has a lower overall dependence than in DR period, as shown in Fig. \ref{fig3}. This indicates that a risk analysis based on linear correlation for MT period is even less appropriate. Actually, at $\alpha=0.01$, all estimated mutual information had significance, while $40.77\%$ and $52.99\%$ of correlations, respectively for DR and MT periods, had no significance, and were set as zero in all further calculations.

\begin{figure}[!ht]
	\begin{center}
		\includegraphics[width=\linewidth]{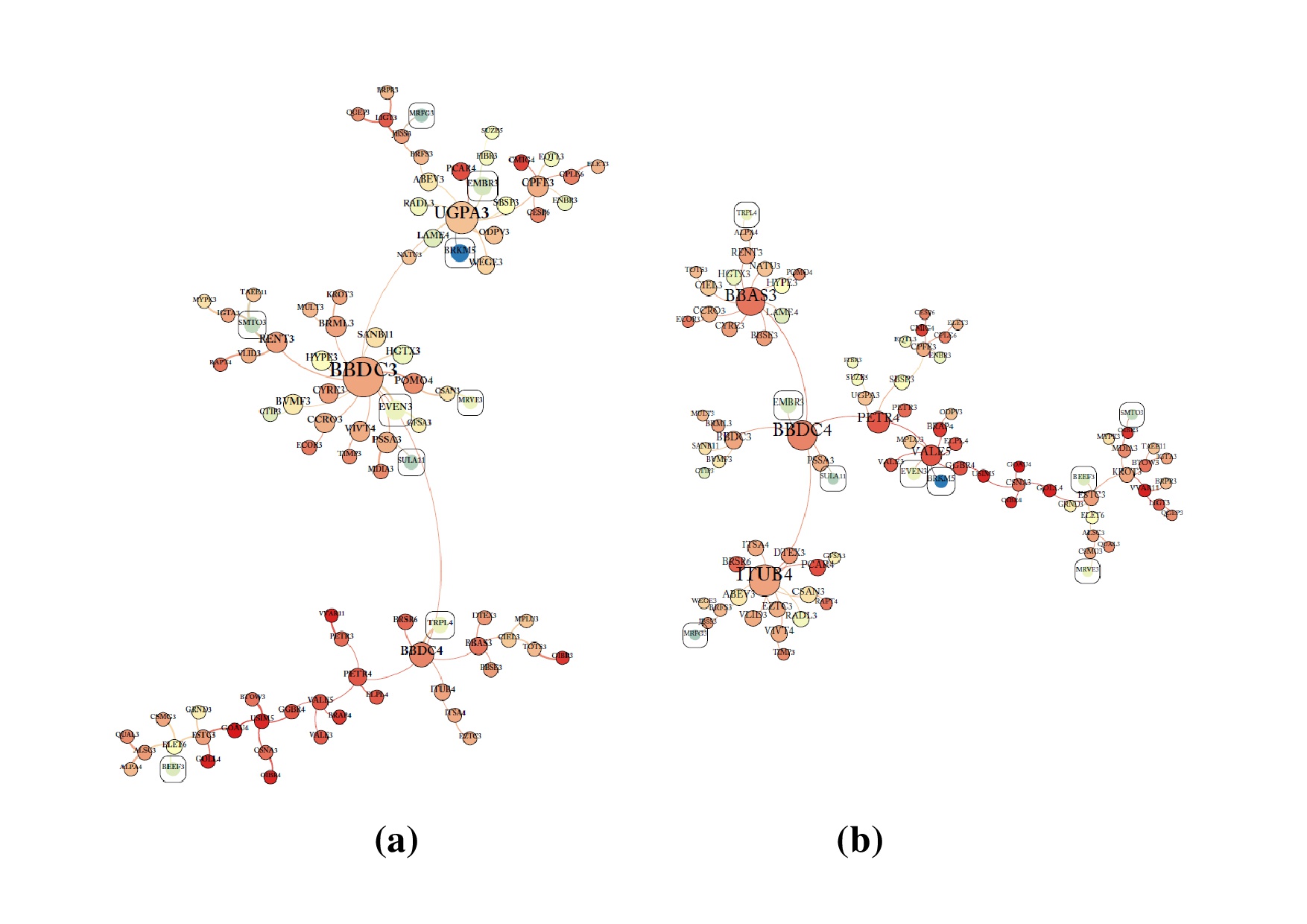}
	\end{center}
	\caption{MST for Dilma Rousseff's period. (a) MST based on linear correlation; (b) MST based on mutual information. Nodes size is related to eigenvector centrality, and color hue degree to the stock variation in the period, red for negative returns, until dark blue, for positive ones. Stocks are highlighted into a square if it composes a minimum of 3\% of the portfolio that maximizes the Sharpe index. MST mean distance: (a) 0.64; (b) 0.56.}
	\label{fig1}
\end{figure}

\begin{figure}[!ht]
	\begin{center}
		\includegraphics[width=\linewidth]{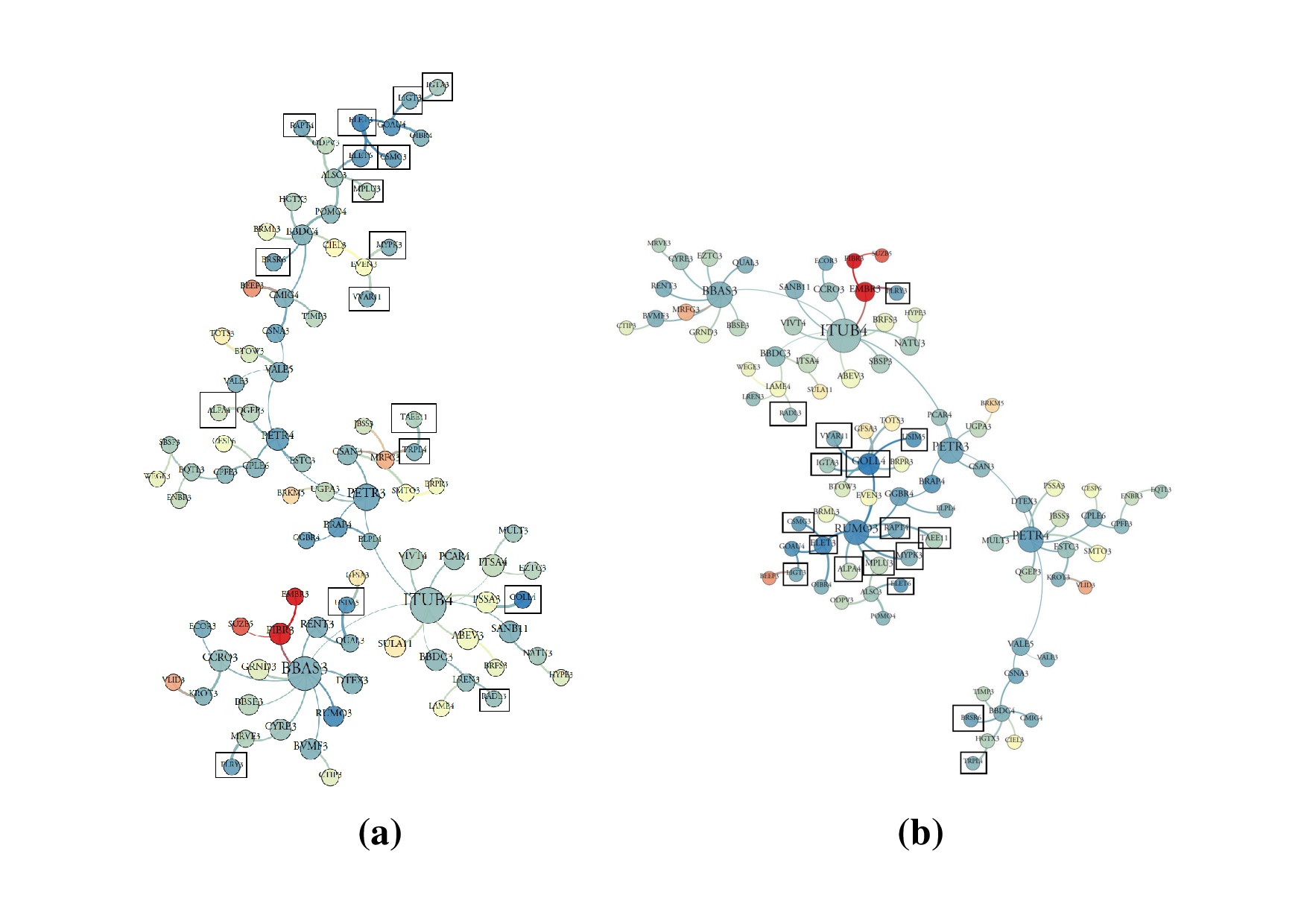}
	\end{center}
	\caption{MST for Michel Temer's period. (a) MST based on linear correlation; (b) MST based on mutual information. Stocks are highlighted into a square if it composes a minimum of 3\% of the portfolio that maximizes the Sharpe index. MST mean distance: (a) 0.66; (b) 0.54.}
	\label{fig2}
\end{figure}

Fig. \ref{fig1}a-b shows the MST network structure for the DR period, obtained from the distances based on linear and global dependence, respectively. To identify individual differences, we highlighted into a square, stocks which may make up the portfolio that maximizes the so-called Sharpe index through the efficient frontier selection technique \cite{sharpe_1964}. This index selects portfolios with the best return-risk ratio.  Fig. \ref{fig2}a-b shows the same as in Fig. \ref{fig1} but for MT period. We observe that MST based on linear correlation the stocks with best Sharpe index are widespread throughout the network for both periods. For the mutual information MSTs, they are concentrated at the periphery for DR period, while they are more central for MT period. The mean distances for each network are depicted in the captions of Figs. \ref{fig1} and  \ref{fig2}, which shows that the  MSTs based on MI are more concentrated, most notably during MT period.

\begin{figure*}[!ht]
\begin{subfigure}{0.48\hsize}\centering
	\includegraphics[width=\linewidth]{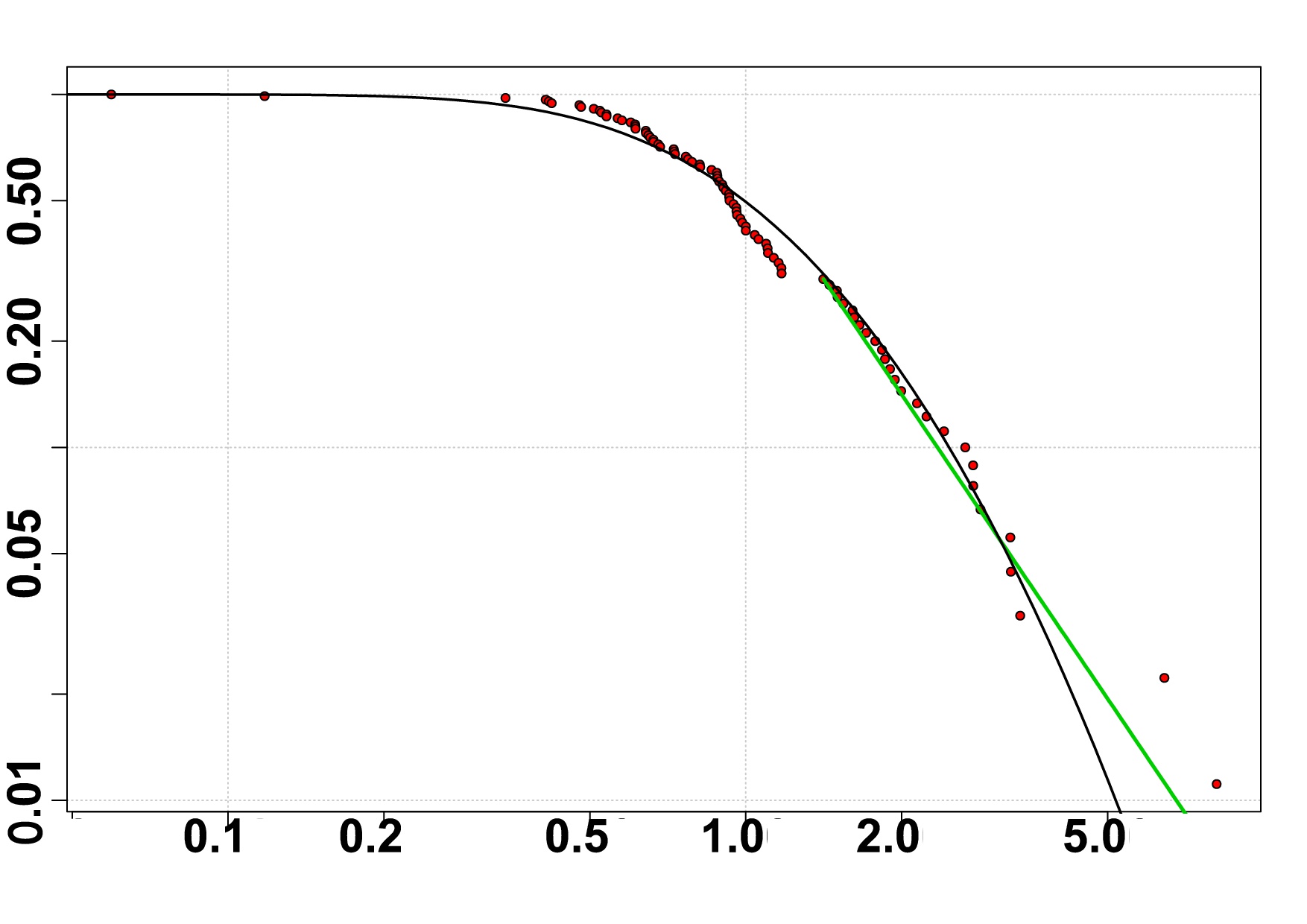}
	\caption{Linear correlation weighted degree distribution for DR period.}
	\label{fig5a}
\end{subfigure}%
\hfill  
\begin{subfigure}{0.48\hsize}\centering
	\includegraphics[width=\linewidth]{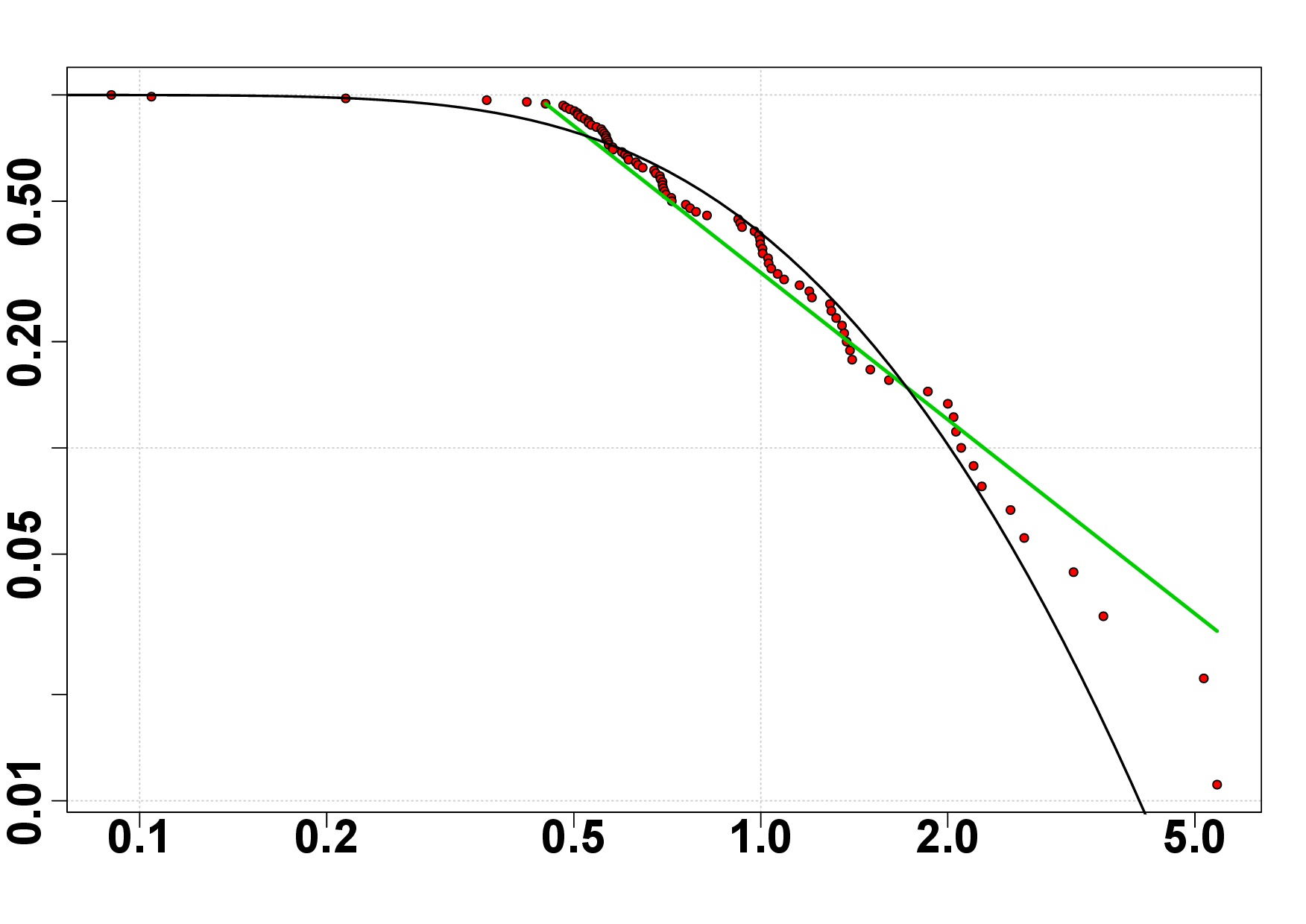}
	\caption{Mutual information weighted degree distribution for DR period.}
	\label{fig5b}
\end{subfigure}
\hfill  
\begin{subfigure}{0.48\hsize}\centering
	\includegraphics[width=\linewidth]{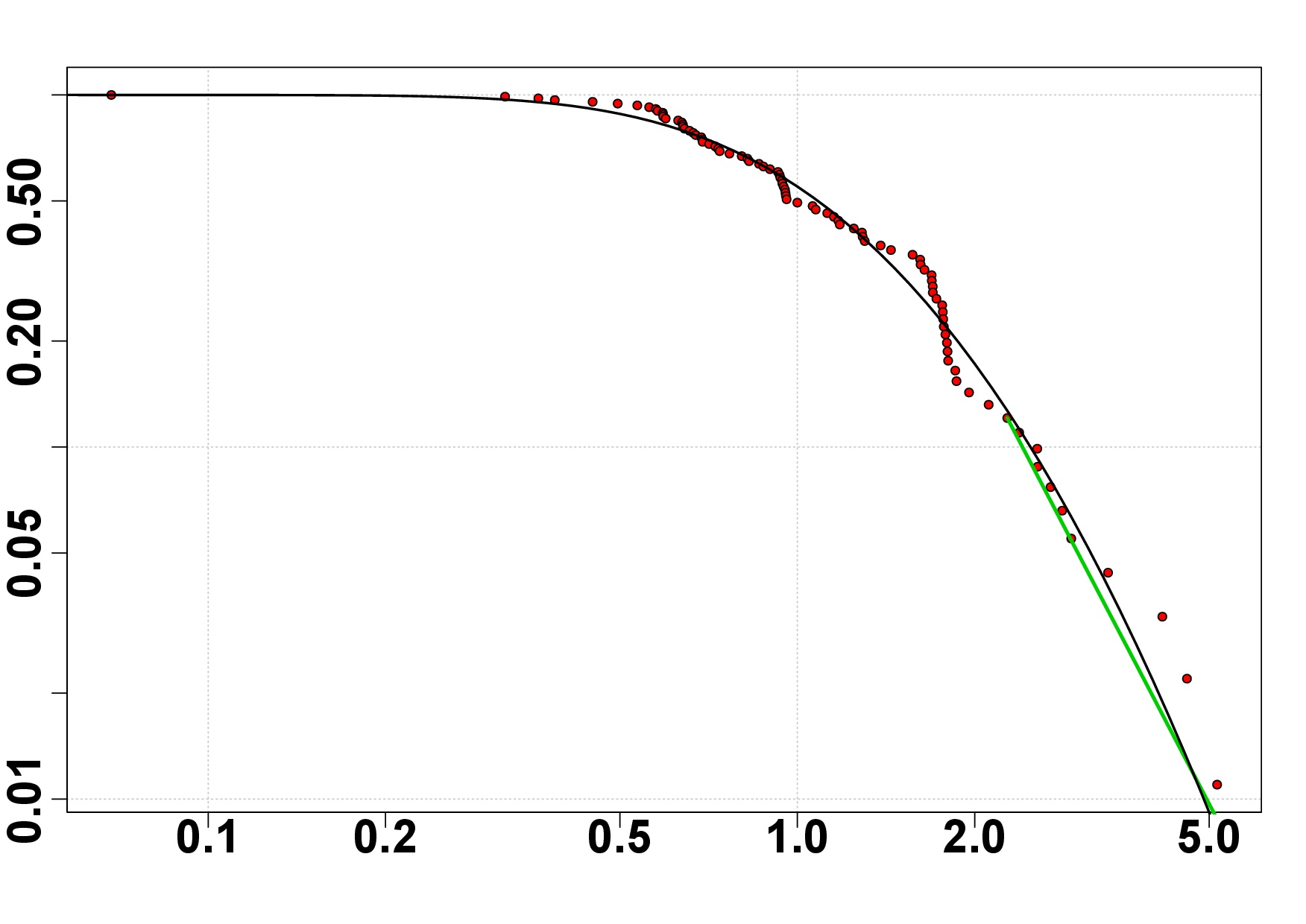}
	\caption{Linear correlation weighted degree distribution for MT period.}
	\label{fig5c}
\end{subfigure}%
\hfill  
\begin{subfigure}{0.48\hsize}\centering
	\includegraphics[width=\linewidth]{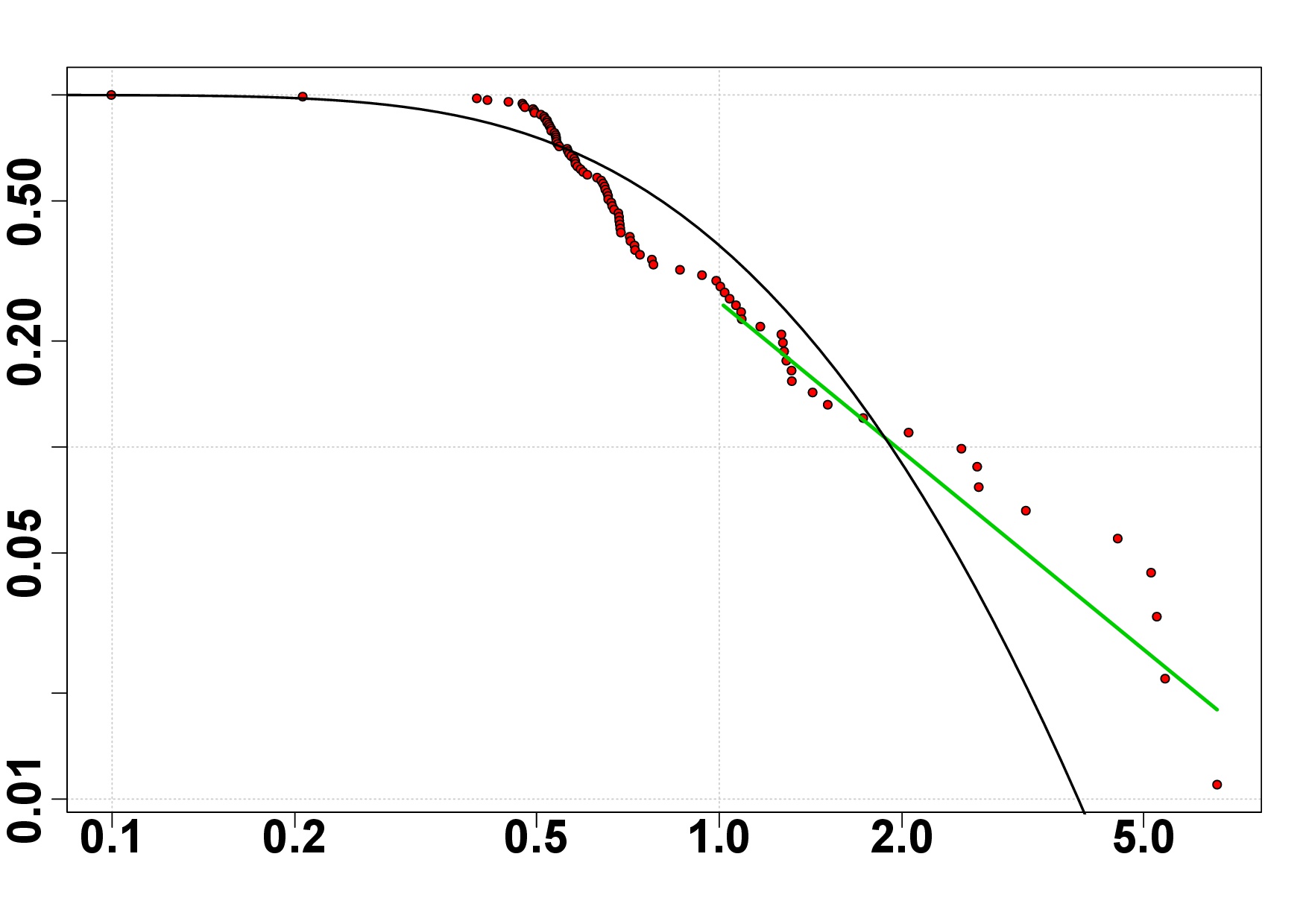}
	\caption{Mutual information weighted degree distribution for MT period.}
	\label{fig5d}
\end{subfigure}
\caption{Weighted degree complementary cumulative distribution function for each period in a log-log scale. The power law fitting is depicted by the green line. Estimated scaling parameter: (a) $\alpha$= 3.08; (b) $\alpha$= 2.44; (c) $\alpha$= 4.19;  (d) $\alpha$= 2.39. The lognormal density fit in black line was included for comparative purposes.}
\label{fig5}
\end{figure*}

Figs. \ref{fig5a}-\ref{fig5d} shows the weighted degree complementary cumulative distribution for the linear correlation network and the one obtained from the global dependence coefficient, highlighted in green the fit to a power law model, for both periods. In order to compare the tails decay we also include a log-normal density fitting (black line). By comparing Figs. \ref{fig5a}-\ref{fig5c} with \ref{fig5b}-\ref{fig5d}, we see that the weighted degree distribution for linear correlation MSTs follows more closely the log-normal distribution, while for the mutual information MSTs the tail dependence is close to a power law. The power law tail is even more evident in Fig. \ref{fig5d}, which corresponds to the period with more nonlinear dependencies. This tail behavior allows that we find nodes with very high degree values with a higher probability.

In Table \ref{tab1}, the assets were splitted into two groups, (a) containing the mean returns of the 20 assets with the highest and lowest eigenvector centrality, and (b) the mean weighted degree of the 20 stocks with the highest and lowest returns for the networks based on linear correlation and global dependence, and where $\mu_{r}$ and $\sigma_r$ stands for mean and standard deviation of returns, and $\mu_{k}$ and $\sigma_k$ are the mean and standard deviation of weighted degree. In this analysis, we are interested in setting the relationship between mean degree structure and stock returns. Table \ref{tab1}(a) shows that, in the MI networks, the most central stocks performed much better than the less central ones for MT period, and exactly the opposite happens for DR period. On the other hand, in the networks based on linear correlation the central and peripheral stocks present similar results. Table \ref{tab1}(b) shows that, in the MI networks, stocks with a very positive performance (return) have a mean weighted degree that is well above the weakest performance group, especially for MT period. That is far beyond what is verified by the network analysis via linear correlation. This indicates that MI networks contains much more useful information about the relationship between centrality and assets returns. For example, as shown in Table \ref{tab1}(a), if an investor, using the information of networks based on mutual information, puts his money on the 20 assets with the largest centralities, during the MT period he would have gained 40\% more than if he had only used an analysis based on linear correlation networks. Furthermore, for the DR period, if the investor uses the information of networks based on mutual information, he would also have saved almost 20\% of loss, providing that he knows the strategy of investing in less central assets. In fact, in Table \ref{tab1}(b), we see that, in the MT period, the best assets returns are related to the highest weighted degrees, and to the smallest in the DR period.

\setlength{\tabcolsep}{0.475em} 
{\renewcommand{\arraystretch}{1.6}
\begin{table}[!htbp]
	\footnotesize{}
	\centering
	\caption{Relationship between centrality and stock returns. (a) Mean returns of the 20 stocks with the highest and lowest eigenvector centrality; (b) weighted degree of the 20 stocks with the highest and lowest returns. $\mu_{r}$ and $\sigma_r$ stands for mean and standard deviation of returns, and $\mu_{k}$ and $\sigma_k$ are the mean and standard deviation of weighted degree. }
	\label{tab1}
	\begin{tabular}{@{}ccccccccc@{}}
		\Xhline{3\arrayrulewidth}
		{ \textit{\textbf{Period}}} & \multicolumn{4}{c}{{\color[HTML]{333333} \textit{Dilma Rousseff}}} & \multicolumn{4}{c}{{\color[HTML]{333333} \textit{Michel Temer}}} \\ \Xhline{3\arrayrulewidth}
		\multicolumn{9}{c}{\textit{\textbf{(a) Returns of the 20 stocks with the highest and lowest eigenvector centralities}}} \\
		\textit{} & \multicolumn{2}{c}{\textit{more central}} & \multicolumn{2}{c}{\textit{less central}} & \multicolumn{2}{c}{\textit{more central}} & \multicolumn{2}{c}{\textit{less central}} \\
		\textit{\textbf{}} & $\mu_r$ & $\sigma_r$ & $\mu_r$ & $\sigma_r$ & $\mu_r$ & $\sigma_r$ & $\mu_r$ & $\sigma_r$ \\
		\textit{Linear correlation} & -20.4\% & 14.7\% & -27.7\% & 33.1\% & 64.3\% & 63.3\% & 64.4\% & 67.9\% \\
		\textit{Global dependence} & -32.7\% & 19.1\% & -8.3\% & 24.9\% & 105.5\% & 126.7\% & 46.5\% & 63.8\% \\ \Xhline{3\arrayrulewidth}
		\multicolumn{9}{c}{\textit{\textbf{(b) Weighted degree of the 20 stocks with the highest and lowest returns}}} \\
		\textit{\textbf{}} & \multicolumn{2}{c}{\textit{best returns}} & \multicolumn{2}{c}{\textit{worst returns}} & \multicolumn{2}{c}{\textit{best returns}} & \multicolumn{2}{c}{\textit{worst returns}} \\
		\textit{\textbf{}} & $\mu_k$ & $\sigma_k$ & $\mu_k$ & $\sigma_k$ & $\mu_k$ & $\sigma_k$ & $\mu_k$ & $\sigma_k$ \\
		\textit{Linear correlation} & 1.45 & 0.83 & 1.90 & 1.21 & 2.2 & 1.4 & 1.5 & 0.76 \\
		\textit{Global dependence} & 1.75 & 1.58 & 2.55 & 2.14 & 3 & 3 & 1.3 & 0.8 \\ \Xhline{3\arrayrulewidth}
	\end{tabular}
\end{table}
}

Table \ref{summary_T} shows the main results of this work. Part (a) of Table \ref{summary_T}  shows some network metrics.  We see that for mutual information networks, the mean MST distance has declined 18\% for period MT and just 12.5\% for period DR. Despite the mean degree remains the same, the robustness coefficient, calculated from Eq. \eqref{mollow3}, increases by 27\% during period MT (during DR, it raises 6.5\%), while power law $\alpha$ parameter, calculated from Eq. \eqref{power}, decreases from 4.19 to 2.39 during period MT (for period DR it decreased from 3.08 to 2.44), which is a significant change. 

The part (b) of Table \ref{summary_T} shows some MSTs qualitative features. As seen in Table \ref{tab1}(b), both linear correlation and mutual information spanning trees suggest that stock returns are related to their centralities, but this is much more evident for networks based on mutual information. In addition, in the linear correlation networks all the best Sharpe index assets are diffuse around the tree. On the other hand, in the mutual information networks, for DR period, the assets with the greatest Sharpe indices appears on the periphery of the graph, while for MT period these same stocks are now much more central. In fact, for DR period, the most central assets 'moved' the market to a 42\% negative return, whereas assets with the best Sharpe index remains into the periphery. During MT period, central assets now push the market to a 50\% positive return. In this period, Sharpe's best indices remain on the central part of the graph. These results suggest that the central stocks are the ones that really 'move' the market.

Finally, Table \ref{summary_T}(c) shows the importance of economic sectors within the networks based on their centrality. We see that Banks sector has the highest centrality in all networks, while industry, construction and energy sectors are always very clustered around network's periphery, so indicating that they are relatively less important for volatility transmission. Indeed, a small number of high linear correlations are sufficient to separate the market sectors, especially banking, the industrial and the electric. In this case, high values of linear correlation also means high values for the global dependence coefficient. Thus, main sector structures did not show relevant changes in networks based on mutual information.

\setlength{\tabcolsep}{0.3em} 
{\renewcommand{\arraystretch}{1.5}
\begin{table}[ht!]
	\footnotesize
	\centering
	\caption{Main results}
	\label{summary_T}
	\begin{tabular}{cclclclcl}
		\Xhline{3\arrayrulewidth}
		{\textit{\textbf{PERIOD}}} & \multicolumn{4}{c}{{\color[HTML]{333333} \textit{\textbf{Dilma Rousseff }}}} & \multicolumn{4}{c}{{\color[HTML]{333333} \textit{\textbf{Michel Temer }}}} \\ 
		\textit{\textbf{DEPENDENCE}} & \multicolumn{2}{c}{\textit{Linear}} & \multicolumn{2}{c}{\textit{Global}} & \multicolumn{2}{c}{\textit{Linear}} & \multicolumn{2}{c}{\textit{Global}} \\ \Xhline{3\arrayrulewidth}
		\multicolumn{5}{c}{\textit{\textbf{(a) MST METRICS}}} \\ 
		\textit{Mean weighted degree} & \multicolumn{2}{c}{1.98} & \multicolumn{2}{c}{1.98} & \multicolumn{2}{c}{1.99} & \multicolumn{2}{c}{1.99} \\ \Xhline{0.5\arrayrulewidth}
		\textit{\begin{tabular}[c]{@{}c@{}}Mean MST  distance\end{tabular}} & \multicolumn{2}{c}{\textit{0.64}} & \multicolumn{2}{c}{\textit{0.56}} & \multicolumn{2}{c}{\textit{0.66}} & \multicolumn{2}{c}{\textit{0.54}} \\ \Xhline{0.5\arrayrulewidth}
		\textit{\begin{tabular}[c]{@{}c@{}}Scale parameter ($\alpha$) \end{tabular}} & \multicolumn{2}{c}{3.08} & \multicolumn{2}{c}{2.44} & \multicolumn{2}{c}{4.19} & \multicolumn{2}{c}{2.39} \\ \Xhline{0.5\arrayrulewidth}
		\textit{\begin{tabular}[c]{@{}c@{}}Robustness coefficient ($R$)\end{tabular}} & \multicolumn{2}{c}{0.61} & \multicolumn{2}{c}{0.65} & \multicolumn{2}{c}{0.56} & \multicolumn{2}{c}{0.71} \\ \Xhline{3\arrayrulewidth}
		\multicolumn{5}{c}{\textit{\textbf{(b) MST MAIN CHARACTERISTICS}}} \\
		\textit{\begin{tabular}[c]{@{}c@{}}Centrality vs. returns\end{tabular}}    & \multicolumn{4}{c}{\begin{tabular}[c]{@{}c@{}}High centrality indicates \\ bad stock returns\end{tabular}} & \multicolumn{4}{c}{\begin{tabular}[c]{@{}c@{}}High centrality indicates \\ good stock returns\end{tabular}} \\ \Xhline{0.5\arrayrulewidth}
		\textit{\begin{tabular}[c]{@{}c@{}}Weighted degree distribution\end{tabular}} & \multicolumn{2}{c}{\begin{tabular}[c]{@{}c@{}}Closer to a\\ log-normal\end{tabular}}  & \multicolumn{2}{c}{\begin{tabular}[c]{@{}c@{}}Power law \\ tail\end{tabular}} & \multicolumn{2}{c}{\begin{tabular}[c]{@{}c@{}}Closer to a \\ log-normal\end{tabular}} & \multicolumn{2}{c}{\begin{tabular}[c]{@{}c@{}}Power law \\ tail\end{tabular}} \\ \Xhline{0.5\arrayrulewidth}
		\textit{\begin{tabular}[c]{@{}c@{}}Sharpe index dispersion \end{tabular}} & \multicolumn{2}{c}{Diffuse} & \multicolumn{2}{c}{Peripheral}  & \multicolumn{2}{c}{Diffuse}     & \multicolumn{2}{c}{Central}  \\ \Xhline{3\arrayrulewidth}
		\multicolumn{5}{c}{\textit{\textbf{(c) ECONOMIC SECTORS IMPORTANCE}}} \\
		\textit{\begin{tabular}[c]{@{}c@{}}Highest centrality \\ (the greatest hub)\end{tabular}} & \multicolumn{2}{c}{\begin{tabular}[c]{@{}c@{}}Banks\\  (Bradesco Bank)\end{tabular}} & \multicolumn{2}{c}{\begin{tabular}[c]{@{}c@{}}Banks\\ (Ita\'u Bank)\end{tabular}} & \multicolumn{2}{c}{\begin{tabular}[c]{@{}c@{}}Banks\\ (Ita\'u Bank)\end{tabular}} & \multicolumn{2}{c}{\begin{tabular}[c]{@{}c@{}}Banks\\ (Ita\'u Bank)\end{tabular}} \\ \Xhline{0.5\arrayrulewidth}
		\textit{\begin{tabular}[c]{@{}c@{}}Lowest centrality \\ \end{tabular}} & \multicolumn{2}{c}{\begin{tabular}[c]{@{}c@{}}Industry and \\ Construction\end{tabular}} & \multicolumn{2}{c}{\begin{tabular}[c]{@{}c@{}}Industry\\ \end{tabular}} & \multicolumn{2}{c}{\begin{tabular}[c]{@{}c@{}}Construction\\ \end{tabular}} & \multicolumn{2}{c}{\begin{tabular}[c]{@{}c@{}}Energy \\ \end{tabular}} \\ 
		\Xhline{3\arrayrulewidth}
	\end{tabular}
\end{table}
}

\section{Conclusion}\label{sec5}
In this paper we studied nonlinear dependencies in Brazilian market network in the period from June/01/2015 to January/26/2016, in which Brazil was under presidency of Dilma Rousseff, and in the period from January/27/2016 to September/08/2016 that includes the government transition from President Dilma Rousseff to President Michel Temer. Minimum Spanning Trees based on mutual information and linear correlation were compared for the two periods. We verified that the mutual information networks bring changes in their structures when compared to the networks based on linear correlation. The asset returns in MT period present a larger number of nonlinear dependencies when compared to the returns of DR period. The network for MT period is the most risky in terms of volatility transmission structure, both by the analysis of the robustness coefficient and by the power law tail evidence in the mutual information minimum spanning trees weighted degree distributions. Finally, mutual information network analysis seems to bring benefits to the investment decision making process, particularly concerning the analysis of network structure and stock performance. Moreover, those type of networks seems to be useful to identify risk not captured by a linear correlation network analysis. For further research it would be interesting to replicate this kind of study for other developing countries or to diverse types of markets, such as currencies; and since our estimated dependencies represents an average effect over each period, also to explore the role played by the time decay of dependencies on the structure of time delayed correlation and mutual information based networks.

Lastly, network analysis is a rapidly developing science, and it would not be surprising if we have much faster, more flexible and interactive methods of data analysis and visualization in a short time. For us, by now it is enough to think that if the key determinant for estimating financial risk models is the dependency structure among all its assets, it seems that having an accurate map at hand is already a great starting point.  

\section{Acknowledgements} 
The authors gratefully acknowledge financial support from Conselho Nacional de Desenvolvimento Científico e Tecnológico (CNPq-Brazil), grant number 134001/2016-8.

\section*{References}
\bibliographystyle{plain}
\bibliography{ref}


\end{document}